\documentclass[11pt]{article}
\usepackage{braket}
\usepackage{graphicx}
\usepackage[fleqn]{amsmath}
\usepackage{color}

\usepackage[top=2.5cm,bottom=2.5cm,left=2.5cm,right=2.5cm]{geometry}
\usepackage{cite}
\begin{document}

\title{An Open-Source Framework for Analyzing $N$-Electron Dynamics: I.\,Multi-Determinantal Wave Functions}

\date{\today}

\author{Vincent Pohl \thanks{corresponding author} \thanks{Institut f\"ur Chemie 
und Biochemie, Freie Universit\"at Berlin, Takustra{\ss}e 3, 14195 Berlin, Germany} 
\thanks{These authors contributed equally to this work.},
Gunter Hermann\footnotemark[2] \footnotemark[3], 
and Jean Christophe Tremblay\footnotemark[2]}

\maketitle

\section*{Abstract}

The aim of the present contribution is to provide a framework for analyzing and visualizing the 
correlated many-electron dynamics of molecular systems, where an explicitly time-dependent electronic wave packet
is represented as a linear combination of $N$-electron wave functions.
The central quantity of interest is the electronic flux density, which contains all information about the transient electronic density, the associated phase, and their
temporal evolution.
It is computed from the associated one-electron operator by reducing the multi-determinantal, many-electron wave packet using the Slater-Condon rules.
Here, we introduce a general tool for post-processing multi-determinant configuration-interaction wave functions obtained at various levels of theory.
It is tailored to extract directly the data from the output of standard quantum chemistry packages using atom-centered Gaussian-type basis functions.
The procedure is implemented in the open-source Python program \textsc{detCI@ORBKIT},
which shares and builds upon the modular design of our recently published post-processing toolbox [{\it J.\,Comput.\,Chem.}~{\bf37}\,(2016)\,1511].
The new procedure is applied to ultrafast charge migration processes in different molecular systems, demonstrating its broad applicability.
Convergence of the $N$-electron dynamics with respect to the electronic structure theory level and basis set size is investigated.
This provides an assessment of the robustness of qualitative and quantitative statements that can be made 
concerning dynamical features observed in charge migration simulations.

\clearpage

\section{Introduction}

The ultrafast evolution of transient electronic densities plays a central role in understanding the chemical reactivity and in predicting spectroscopic properties of molecules.
With the recent advances in attosecond laser technologies, it has now become possible to indirectly observe the dynamics of electrons on
their natural timescale \cite{01:Krausz,02:Krausz,02:Krausz:innershell,03:B:atto,03:Krausz,04:Villeneuve,05:Wurth,06:Vrakking,07:Krausz,09:Krausz,10:Woerner}.
Whereas direct observation of the electron flow remains elusive, its experimental reconstruction yields a wealth of information about charge migration in molecules,
opening a wide range of new applications.

The theoretical description of electron dynamics has also greatly progressed over the last decades
\cite{87:K:tdhf,93:GE,95:PGG,97:YB,99:RL,00:CRSU,00:HKKF,03:BC,03:SM,03:K:tdci,03:LB,03:ZKFBS,04:KK:mctdhf,04:CC,05:P,05:BMG,
nestel1,05:KKS:licn,07:KKS:tdci6,06:BM,06:BMSY,06:K:quino,06:rubio,schlegel07,08:SBN,08:SGY,08:TKS,tdci3,tdci4,09:KN,
09:N,11:TKKS,13:KT:bmq7h,13:madsen,14:madsen:miyagi,14:madsen}.
In particular, Time-Dependent Density Functional Theory  (TDDFT) \cite{84:rg:tddft} holds lot of promises due to its computational efficiency and intuitive interpretation.
Approaches based on $N$-electron wave functions such as Multi-Configuration Time-Dependent Hartree-Fock \cite{03:ZKFBS,04:KK:mctdhf,nestel1}
or time-dependent Configuration Interaction (CI) \cite{03:K:tdci,huber2ppe,krause05-1,05:KKS:licn,08:KK:tdci,tdci3,tdci4,11:Kroener,13:RRN, 14:Dutoi}
offer an attractive alternative to density-based schemes. 
They share the common philosophy of representing an $N$-electron wave packet as a linear combination of spin-symmetrized Slater-determinants, which 
are constructed by exciting electrons from a single reference determinant.
As such, the time-evolving wave packet is thus a multi-determinantal wave function.
These methods are systematically improvable and converge to the same Full CI limit.
Their major limitation is the high associated numerical cost, but this problem is mitigated by the ever increasing computational resources at our disposal.

The choice of an $N$-electron determinant basis for the electron dynamics has the important advantage of ensuring the $N$-representability
of the wave packet at all times. It also reveals information about the dynamical build up of correlation in the transient electronic wave packet and its
physical origin (particle-hole ($p$--$h$), two-particle--two-hole (2$p$--2$h$), $\dots$, excitations).
On the other hand, it has the associated disadvantage of rendering interpretation of the electron dynamics less intuitive, which can be circumvented by 
using a proper set of visualization tools. Apart from the transient one-electron density, these need to include information about the evolution of the phase
of the wave packet, which strongly depends on the electronic correlations. This complementary information is encoded in the electronic flux density,
equivalent to the current density, from which a qualitative picture of the electron flow  emerges naturally. The electronic flux density is a vector field that allows,
at first glance, for a microscopic understanding of the mechanisms at work during charge migration processes. 

In this contribution, we introduce a framework for analyzing and visualizing the correlated many-electron dynamics of molecular systems based on 
the reconstruction of the electronic flux density from a general multi-determinantal wave function. This requires a number of 
fundamental one-electron quantities, such as difference electronic densities, transient electronic flux densities, and transition dipole moments, 
that are not directly accessible from the output of standard quantum chemistry packages.  
Our initiative embraces the open-source molecular modelling philosophy  \cite{Pirhadi2016127} and 
builds upon the modular structure of our recently published quantum chemistry toolbox \hbox{\textsc{ORBKIT}} \cite{orbkit}.
The latter is capable of computing a multitude of static electronic properties based on
the data of electronic structure calculations from single-determinant wave function 
approaches, such as Hartree-Fock (HF) or Density Functional Theory (DFT) methods.
Here, we extend the capabilities of \textsc{ORBKIT} to multi-determinant wave functions
by exploiting its highly modular and easily comprehensible Python architecture.
The new customized post-processing program, \textsc{detCI@ORBKIT},
can extract the data of multi-determinantal wave functions from various quantum 
chemistry programs using atom-centered Gaussian-type basis functions, and evaluate matrix elements
of one-electron operators in the basis of $N$-electron eigenstates.
The time-dependent quantities required for analyzing and visualizing the $N$-electron wave packet dynamics
by means of the flux density are then calculated as linear combinations of the static matrix elements.
This new tool will prove valuable to investigate a great number of charge migration processes.
The present contribution also explores parameters that affect the qualitative
and quantitative analysis of the electronic flux density, as applied to the 
electron dynamics in H$_3^+$ and LiH.

The paper is organized as follows: Section \ref{theory} briefly describes the time-dependent
configuration-interaction methodology and introduces general computational rules for computing
one-electron matrix elements. The influence of the basis set size on the flux density is benchmarked in
subsection \ref{h3+} for the H$_3^+$ test system. Subsection \ref{lih} investigates the influence of the 
electronic structure method on the qualitative features of the electron migration process. 
Concluding remarks are presented in Section \ref{concl}.
Atomic units are used throughout the paper ($\hbar = m_e = e = 4\pi\varepsilon_0 = 1$), unless stated otherwise.

\section{Computational Procedure and Theory}\label{theory}

\subsection{Time-Dependent Configuration Interaction}

The electron dynamics of a molecular system can be described by solving the non-relativistic 
time-dependent electronic Schr{\"o}dinger equation\cite{Schroedinger1926quantisierung}
\begin{eqnarray}\label{TDSE}
i\frac{\partial}{\partial t} \Ket{\Psi_{{\rm el}}\left( t\right)} = \hat{H} \Ket{\Psi_{{\rm el}}\left( t\right)} .
\end{eqnarray}
The field-free Hamiltonian $\hat{H}$ for a system consisting of $N$ electrons and
$N_A$ nuclei is written in the clamped nuclei approximation as
\begin{eqnarray}\label{hamiltonian}
    \hat{H} = - \frac{1}{2} \sum\limits_{i=1}^{N} \nabla_i^2 + \sum\limits_{i=1}^{N} \sum\limits_{j>i}^{N} \frac{1}{r_{ij}} - \sum\limits_{i=1}^{N} 
\sum\limits_{A=1}^{N_A} \frac{Z_A}{r_{Ai}} \quad \text{.}
\end{eqnarray}
Here, $r_{ij}$ is the distance between electrons $i$ and $j$, $Z_A$ is the charge number of nucleus $A$, and $r_{Ai}$ is the distance between
nucleus $A$ and  electron $i$.
In general, the multi-particle time-dependent electronic wave function $\Ket{\Psi_{{\rm el}}\left( t\right)}$ 
can be formulated as a linear superposition of stationary electronic states $\Ket{\Phi_\lambda}$
\begin{eqnarray}\label{td_wf}
\Ket{\Psi_{{\rm el}}\left(t\right)} = \sum_{\lambda} B_{\lambda}\left( t\right) \Ket{\Phi_\lambda}
\end{eqnarray}
with $B_{\lambda}\left( t\right)$ as the time-dependent expansion coefficients of state $\lambda$.
From a dynamical perspective, it is convenient to choose a basis of  $N$-electron states that diagonalizes
the field-free Hamiltonian at a given level of theory. This is the approach followed in the present paper.

Generally speaking, the time-independent $N$-electron wave function $\Ket{\Phi_\lambda}$
can be expressed in the terms of a configuration interaction (CI) expansion. 
CI methodologies are conceptually similar to other high-level wave function based methods, as will be discussed below.
To ensure a proper description of electron correlation, 
the wave function is expanded in a complete set of configuration state functions\cite{szabo}
\begin{eqnarray}\label{general_CI}
 \Ket{\Phi_{\lambda}^{{\rm CI}}} = \sum_{p} C_{p}^{\left(\lambda\right)} \Ket{\phi_{p}},
\end{eqnarray}
where  the expansion parameters (or CI-coefficients) $C_{p}^{\left(\lambda\right)}$ are optimized variationally.
In the present implementation, the orthonormal configurations $\Ket{\phi_{p}}$ are chosen as Slater determinants.
These are defined as antisymmetrized products of one-electron spin orbitals $\Ket{\varphi_a}$
\begin{eqnarray}
\phi\left(\mathbf{x}_{1},\,\mathbf{x}_{2},\,\dots,\,\mathbf{x}_{N}\right) & = & \frac{1}{\sqrt{N!}}\left|\begin{array}{cccc}
\varphi_{a}\left(\mathbf{x}_{1}\right) & \varphi_{b}\left(\mathbf{x}_{2}\right) & \ldots & \varphi_{c}\left(\mathbf{x}_{N}\right)\\
\varphi_{a}\left(\mathbf{x}_{1}\right) & \varphi_{b}\left(\mathbf{x}_{2}\right) & \ldots & \varphi_{c}\left(\mathbf{x}_{N}\right)\\
\vdots & \vdots & \ddots & \vdots\\
\varphi_{a}\left(\mathbf{x}_{1}\right) & \varphi_{b}\left(\mathbf{x}_{2}\right) & \ldots & \varphi_{c}\left(\mathbf{x}_{N}\right)
\end{array}\right| \label{sd}\\
& \equiv & \Ket{\varphi_{a}\varphi_{b}...\varphi_{c}}. \label{sh_sd}
\end{eqnarray}
Here, the Slater determinant is represented as a function of the spin and spatial 
coordinates $\mathbf{x}$ of the $N$ electrons,
and $\left\lbrace \varphi_{a}\varphi_{b}...\varphi_{c} \right\rbrace$ are the occupied orthonormal molecular spin orbitals. 
In the CI-approach, the various Slater determinants are constructed by excitations 
of spin orbitals from a single reference state $\Ket{\phi_{0}}$.
For example, exciting an electron from an occupied spin orbital $a$ to a 
virtual spin orbital $r$ from the reference state forms a singly excited determinant 
$\Ket{\phi_{a}^{r}}$.
Accordingly, the CI wave function can be reformulated in terms of these excited determinants
\begin{eqnarray}\label{CI}
\Ket{\Phi_{\lambda}^{{\rm CI}}} = C_{0}^{\left(\lambda\right)} \Ket{\phi_{0}}+\sum_{ar}C_{a}^{r\left(\lambda\right)}\Ket{\phi_{a}^{r}}
+\sum_{abrs}C_{ab}^{rs\left(\lambda\right)}\Ket{\phi_{ab}^{rs}} + \ldots\,,
\end{eqnarray}
where $\left\lbrace a,b,c \right\rbrace$ are the occupied spin orbitals, and $\left\lbrace r,s,t \right\rbrace$
denote the virtual spin orbitals (i.e., unoccupied in the reference determinant).
The exact ansatz including all possible excitations is referred to as Full CI approach.
Since the number of conceivable excitations increases factorially with the number of electrons 
and orbitals, this is computationally very expensive and only feasible for small systems.
To circumvent this bottleneck, two main strategies are pursued to reduce the Full CI expansion (cf. Eq.~\eqref{CI}): 
first, the truncation to a certain maximum rank of excitations, e.g., CI Singles (CIS) or CI Doubles (CID),
and second, the restriction of the active space to a certain number of 
electrons in a specified number of orbitals, e.g., Multi-Configuration Self 
Consistent Field (MCSCF)\cite{ruud1994multiconfigurational}.
In the latter scheme, the orbitals themselves appearing in Eq.~\eqref{sh_sd} are variationally optimized
in addition to the CI-coefficients. This yields a better representation of the correlation in a reduced orbital space,
usually brought to the Full CI limit in an active space chosen close the HOMO-LUMO gap of the molecule. 
Quite importantly, all information required to reconstruct the orbitals and the $N$-electron eigenfunctions
at a chosen level of theory are accessible from the output of standard quantum chemistry packages.
Indeed, this information is used by many post-processing programs for computing orbital-derived quantities.
We will now investigate how it is possible to use this knowledge to compute time-dependent wave function
derived properties.

\subsection{General Considerations on Expectation Values}
Exploiting the structure of the time-dependent multi-determinant wave functions, Eqs.~\eqref{td_wf} and \eqref{general_CI},
the expectation value of any one-electron operator $\hat{F}$  can be expressed as
\begin{eqnarray}
\Braket{\hat{F}}(t) & = & \Braket{\Psi_{{\rm el}}\left(t\right) \left| \hat{F} \right| \Psi_{{\rm el}}\left(t\right)} \label{exp_val_1} \\
& = & \sum_{\lambda\nu} B_{\lambda}^{\dagger}\left( t\right) B_{\nu}\left( t\right) \Braket{\Phi_{\lambda}^{{\rm CI}} \left| \hat{F} \right| \Phi_{\nu}^{{\rm CI}}} \label{exp_val_2}\\
& = & \sum_{\lambda\nu} B_{\lambda}^{\dagger}\left( t\right) B_{\nu}\left( t\right) \sum_{pq} C_{p}^{\left(\lambda\right)} C_{q}^{\left(\nu\right)} \Braket{\phi_{p} \left| \hat{F} \right| \phi_{q}} \label{exp_val_3}.
\end{eqnarray}
In order to obtain the expectation value of the operator $\hat{F}$, one has to evaluate 
the matrix elements $\Braket{\phi_{p} \left| \hat{F} \right| \phi_{q}}$ between two determinants (cf. Eq.~\eqref{exp_val_3}).
For that purpose, the Slater-Condon rules\cite{slater1929theory,condon1930theory,slater1931theory},
allow to express the respective matrix elements in terms of one-electron integrals in the spin orbital space.
These rules can be summarized as follows
$\Braket{\phi_{p} \left| \hat{F} \right| \phi_{q}}$ for three general types:
\begin{enumerate}
  \item Identical Slater determinants
  \begin{eqnarray}\label{scr_id}
  \Braket{\cdots abc \cdots \left| \hat{F}\right| \cdots abc \cdots} & = & \Braket{\cdots ab \cdots \left|\hat{F} \right| \cdots ab \cdots}\nonumber \\
  & = & \sum_{a} \Braket{ \varphi_a \left| \hat{F}\right| \varphi_a}.
  \end{eqnarray}
  \item Two Slater determinants differing by a single spin orbital
  \begin{eqnarray}\label{scr_one}
  \Braket{\cdots rbc \cdots \left| \hat{F}\right| \cdots abc\cdots}  & = & \Braket{\cdots rb \cdots \left|\hat{F} \right| \cdots ab \cdots}\nonumber \\
  & = & \Braket{ \varphi_r \left| \hat{F}\right| \varphi_a }.
  \end{eqnarray}
  \item Two determinants which differ by two or more spin orbital
  \begin{eqnarray}\label{scr_two}
  \Braket{\cdots ars \cdots \left| \hat{F}\right| \cdots abc\cdots}  =  0.
  \end{eqnarray}
\end{enumerate}
A prerequisite for applying the Slater-Condon rules is the maximum coincidence principle,
i.e., all common spin orbitals of both configurations appear at the same positions in the respective Slater determinant.
This is achieved by permutation of the spin orbitals in one of the determinants, i.e., by interchanging columns in Eq.~\eqref{sd}.
This necessary re-ordering can change the sign due to the antisymmetric properties of determinants
\begin{eqnarray}
  \Ket{\phi} = \Ket{\cdots abc\cdots} = 
  - \Ket{\cdots acb \cdots} = \Ket{\cdots cba \cdots}.
\end{eqnarray}
The resulting phase factor, $\left(-1\right)^{N_{{\rm P}}}$, can be determined by counting the 
required number of column permutations $N_{{\rm P}}$ to reach maximum coincidence.
It must be accounted for when applying the Slater-Condon rules to compute the matrix elements in Eq.~\eqref{exp_val_3}.

The computation of expectation values of any one-electron operator from a configuration interaction wave packet of the 
type Eq.~\eqref{general_CI} boils down to evaluate transition moments between spin orbitals.
In computational chemistry, the spin orbitals are usually transformed to 
spin-free representations by integrating over the spin coordinates.
We specialize here to the case, where these spatial molecular orbitals (MO) are defined 
in the framework of the MO-LCAO (Molecular Orbital - Linear Combination of Atomic Orbitals) ansatz.
Specifically, an  MO $\varphi_{a}$ is expanded using a finite set of atom-centered basis functions
\begin{eqnarray}\label{molcao}
\varphi_{a} \left(\mathbf{r}\right) = \sum_{A=1}^{N_{A}}\sum_{i_A=1}^{n_{{\rm AO}}{(A)}} D^{(a)}_{i_A} \chi_{i_A}\left(\mathbf{r}-\mathbf{R}_{A}\right),
\end{eqnarray}
with $D^{(a)}_{i_A}$ as the $i_A$th expansion coefficient for MO $a$.
The basis functions $\chi_{i_A}$ are atomic orbitals, expressed as a function of the Cartesian coordinates of one electron $\mathbf{r}$
and the spatial coordinates $\mathbf{R}_{A}$ of nucleus $A$.
$N_{{A}}$ labels the number of atoms and $n_{{\rm AO}}{(A)}$ denotes the number of atomic orbitals on atom $A$,
with  $N_{\rm AO}=\sum_{A=1}^{N_A}n_{{\rm AO}}{(A)}$.
In the MO-LCAO representation, the transition moments between spin orbitals take the form
\begin{eqnarray}
\Braket{\phi_{p} \left| \hat{F} \right| \phi_{q}} =
 \sum_{A,B}^{N_{{A}}}\sum_{i_A=1}^{n_{{\rm AO}}{(A)}}\sum_{j_B=1}^{n_{{\rm AO}}{(B)}} D^{(p)}_{i_A} D^{(q)}_{j_B} \Braket{\chi_{i_A} \left| \hat{F} \right| \chi_{j_B}}
\end{eqnarray}
All required information to reconstruct molecular orbitals, i.e., the MO-LCAO coefficients $D^{(a)}_{iA}$ and the 
atom-centered basis functions, can be found in the output of standard quantum chemistry program packages.
The integrals in the atomic orbital basis are computed analytically with our Python post-processing toolbox \textsc{ORBKIT}
for a wide range of one-electron operators \cite{orbkit}.

\subsection{The Electronic Continuity Equation}
A widespread quantity for the visual representation of electronic motions in molecular systems is the time-dependent one-electron density 
$\rho \left(\mathbf{r},t \right)$.\cite{nafie1997electron,Freedman1998,hermann2014electronic,Hermann2016}
It remains a useful tool to characterize the correlated electron dynamics from multi-determinant wave functions of the form Eq.~\eqref{general_CI},
and it can be computed as an expectation value of the associated one-electron density operator
\begin{eqnarray}\label{rho_op}
\hat{\rho} \left( \mathbf{r}\right) = \sum_{k}^{N} \delta\left( \mathbf{r} - \mathbf{r}_{k} \right)= \sum_{k}^{N} \delta_k(\mathbf{r}),
\end{eqnarray}
where $\delta\left( \mathbf{r} - \mathbf{r}_{k} \right)= \delta_k(\mathbf{r})$ is the Dirac delta distribution, $\mathbf{r}$ designates the grid of observation, and
$\mathbf{r}_{k}$ refers to the position of electron $k$.
The one-electron density admits a realistic interpretation as a probability fluid, which must satisfy a continuity equation of the form
\begin{eqnarray}\label{con_eq}
\frac{\partial}{\partial t} \rho \left(\mathbf{r},t \right) = - \vec{\nabla} \cdot \mathbf{j} \left(\mathbf{r},t\right).
\end{eqnarray}
The vector field $\mathbf{j} \left(\mathbf{r},t\right)$ is the electronic flux density, often called current density, 
which contains information about electronic phase driving the spatial flow of the electron density.
The associated operator can be written as
\begin{eqnarray}\label{j_op}
\hat{j} \left( \mathbf{r}\right) = \frac{1}{2}\sum_{k}^{N} \left(\delta_k(\mathbf{r})\hat{p}_{k} + \hat{p}^{\dagger}_{k}\delta_k(\mathbf{r})\right).
\end{eqnarray}
Here, $\hat{p}_{k}=-i\vec{\nabla}_{k}$ is the momentum operator of an electron  $k$,
where $\vec{\nabla}_{k}$ is the gradient operator.

Using the time-dependent wave function of Eq.~\eqref{td_wf}, the expectation value for the electron density reads
\begin{eqnarray}\label{rho}
\rho \left(\mathbf{r},t \right) & = &  \int\Psi_{\rm{el}}^{\dagger}\left(\mathbf{r}^N, t\right) \hat{\rho}\left(\mathbf{r}\right) \Psi_{\rm{el}} \left(\mathbf{r}^N, t\right) {\rm{d}}\mathbf{r}^{N}\\
& = & \sum_{\lambda\nu} B_{\lambda}^{\dagger}\left(t\right)B_{\nu}\left(t\right) \int\left(\Phi_{\lambda}^{\rm{CI}}\left(\mathbf{r}^N\right)\hat{\rho}\left(\mathbf{r}\right) \Phi_{\nu}^{\rm{CI}}\left(\mathbf{r}^N\right)\right) {\rm{d}}\mathbf{r}^{N}.
\label{rho_matrix}
\end{eqnarray}
with $\Psi_{{\rm el}}$ as a function of the spatial coordinates $\mathbf{r}^N$ of $N$ electrons.
In Eq.~\eqref{rho_matrix}, the matrix expression can be simplified by applying the 
Slater-Condon rules
\begin{eqnarray}\label{rho_CI_mo}
\int\Phi_{\lambda}^{\rm{CI}}\left(\mathbf{r}^N\right)\hat{\rho}\left(\mathbf{r}\right) \Phi_{\nu}^{\rm{CI}}\left(\mathbf{r}^N\right){\rm{d}}\mathbf{r}^{N}& = & 
\sum_{p} C_{p}^{\left( \lambda \right)} C_{p}^{\left( \nu \right)}
\sum_{\overline{a}} n_{\overline{a}} \left|  \varphi_{\overline{a}}(\mathbf{r}) \right|^2 \nonumber\\
&&+ \sum_{p \neq q} C_{p}^{\left( \lambda \right)} C_{q}^{\left( \nu \right)}\left(\varphi_{\overline{r}}(\mathbf{r})\varphi_{\overline{a}}(\mathbf{r})\right)
\end{eqnarray}
with $n_{\overline{a}}$ as the occupation number of MO $\overline{a}$. The over-line ``--'' denotes a formal excitation from the MO $\overline{a}$ to MO $\overline{r}$ taking the configuration state function
of state $\nu$ as a reference.
The corresponding expression for the electron flux density can be formulated as
\begin{eqnarray}\label{j}
\mathbf{j} \left(\mathbf{r},t\right) & = & \int\Psi^{\dagger}_{\rm{el}}\left(\mathbf{r}^N, t\right) \hat{j}(\mathbf{r}) \Psi_{\rm{el}}\left(\mathbf{r}^N, t\right) {\rm{d}}\mathbf{r}^{N}\\
 & = &  \sum_{\lambda\,\nu}
B^{\dagger}_{\lambda}\left(t\right)B_{\nu}\left(t\right) \int\left(\Phi_{\lambda}^{\rm{CI}}\left(\mathbf{r}^N\right)\hat{j}(\mathbf{r}) \Phi_{\nu}^{\rm{CI}}\left(\mathbf{r}^N\right)\right) {\rm{d}}\mathbf{r}^{N}\nonumber\\
& = & 2 i\sum_{\lambda<\nu}
{\rm Im}\left[B^{\dagger}_{\lambda}\left(t\right)B_{\nu}\left(t\right)\right] \mathbf{j}_{\lambda \nu} \left(\mathbf{r},t\right) \label{j_matrix} ,
\end{eqnarray}
where $\mathbf{j}_{\lambda \nu} \left(\mathbf{r},t\right)$ is the transition electronic flux density from state $\lambda$ to state $\nu$
\begin{eqnarray}
\mathbf{j}_{\lambda \nu} \left(\mathbf{r},t\right) = -\frac{i}{2} \sum\limits_k\left(
\int\delta_{k} (\mathbf{r})\left(\Phi_{\lambda}^{\rm{CI}}\left(\mathbf{r}^N\right) \vec{\nabla}_k \Phi_{\nu}^{\rm{CI}}\left(\mathbf{r}^N\right)\right) {\rm{d}}\mathbf{r}^{N}\right.\nonumber\\
\left.-\int\delta_{k} (\mathbf{r})\left(\Phi_{\nu}^{\rm{CI}}\left(\mathbf{r}^N\right) \vec{\nabla}_k\Phi_{\lambda}^{\rm{CI}}\left(\mathbf{r}^N\right)\right) {\rm{d}}\mathbf{r}^{N}
\right)\label{transitionJ}.
\end{eqnarray}
Since the electronic states are real-valued, the diagonal terms $\mathbf{j}_{\lambda\lambda} \left(\mathbf{r},t\right)$, i.e., the adiabatic flux density
\cite{diestler2011cct,diestler2011cc_appl,bredtmann2015quantum,pohl2016adiabatic,Schild2016}, vanish.
The same argument holds for all matrix elements in Eq.~\eqref{j_matrix} involving identical determinants.
Using the Slater-Condon rules, the integrals in Eq.~\eqref{transitionJ}  simplify to one-electron integrals over the spin orbitals
\begin{eqnarray}\label{j_CI_mo}
\sum_{k} \int\delta_{k} (\mathbf{r})\left(\Phi_{\lambda}^{\rm{CI}}\left(\mathbf{r}^N\right)\vec{\nabla}_k\Phi_{\nu}^{\rm{CI}}\left(\mathbf{r}^N\right)\right) {\rm{d}}\mathbf{r}^{N} = 
\sum_{p \neq q} C_{p}^{\left( \lambda \right)} C_{q}^{\left( \nu \right)}\left(\varphi_{\overline{r}}(\mathbf{r})\vec{\nabla}\varphi_{\overline{a}}(\mathbf{r}) \right).
\end{eqnarray}
The derivatives of the molecular orbitals are computed analytically using functions from the Python toolbox \textsc{ORBKIT},
with which both, the density and the electronic flux density, can then be projected on an arbitrary grid.

From the one-electron density, it is possible to derive another potentially useful quantity for the analysis of $N$-electron dynamics.
The difference density $\mathbf{y} \left(\mathbf{r},t \right)$ describes the variation of the electron density within the time interval 
$\left[0,t\right]$ from a chosen initial condition\cite{barth2009concerted,Berg}.
It is defined as the integral over time of the electron flow
\begin{eqnarray}\begin{aligned}\label{yield}
  \mathbf{y} \left(\mathbf{r},t \right) &= \int_{0}^{t} {\rm d} t' \frac{\partial \rho \left(\mathbf{r},t' \right)}{\partial t'} = \rho \left(\mathbf{r},t \right) - \rho \left(\mathbf{r},0 \right)
\end{aligned}\end{eqnarray}
The difference density determines the number of electrons that have moved in and out 
of a specific volume element during a given laps of time. As such, it yields quantitative information that is complementary to both the 
electronic flux density and the electron flow.

The convergence of the continuity equation can be {\sl a priori} estimated by inspecting closely related static quantities
that relates expectation values of $\hat{\rho} \left( \mathbf{r}\right)$ and $\hat{\jmath} \left( \mathbf{r}\right)$. 
A potentially useful tool in this endeavor is the comparison of the dipole operator in length and in velocity gauge.\cite{nafie1997electron}
The former derives from the one-electron density and takes the form
\begin{eqnarray}\label{mu_op_lg}
\hat{\mu}_{r} = - \int \mathbf{r} \hat{\rho} \left( \mathbf{r}\right) {\rm d}\mathbf{r}.
\end{eqnarray} 
The latter is defined from the electronic flux density as
\begin{eqnarray}\label{mu_op_vg}
\hat{\mu}_{v} = - \int \hat{\jmath} \left( \mathbf{r}\right) {\rm d}\mathbf{r}.
\end{eqnarray}
As the transition moment between a given pair of states $\{\Phi_{\lambda},\Phi_{\nu}\}$, both can be directly related to each other via\cite{bransden1983physics}
\begin{eqnarray}\label{mu_vg_to_lg}
\Braket{\left(\hat{\mu}_v\right)_r}_{\lambda\nu} &=& \frac{i}{\left( E_{\nu} - E_{\lambda}\right)} \Braket{\Phi_{\lambda}^{{\rm CI}}| \hat{\mu}_{v}| \Phi_{\nu}^{{\rm CI}}}.
\end{eqnarray}
The last expression can be used to estimate the quality of the level of theory 
and of the underlying basis of a given quantum chemical calculation.
For a converged calculation, the transition moment $\Braket{\hat{\mu_{r}}}_{\lambda\nu}$ (Eq.~\eqref{mu_op_lg}) and 
$\Braket{\left(\hat{\mu}_v\right)_r}_{\lambda\nu}$ (Eqs.~\eqref{mu_op_vg} and \eqref{mu_vg_to_lg}) must become identical
for every pair of states involved in the dynamics.

\subsection{Implementation Details}

The computational rules described in the previous subsections are implemented in our new open-source Python toolbox \textsc{detCI@ORBKIT},
which is freely available at \linebreak\texttt{https://github.com/orbkit/orbkit/}.
The program requires a preliminary quantum chemistry calculation at a desired level of wave function theory, and using 
Gaussian-type atom-centered orbitals. Starting from the data of a determinantal CI-calculation, the program builds a library
of transition moments and expectation values of one-electron operators projected on an arbitrary grid,
to be used for analyzing the $N$-electron dynamics.
The electronic ground state and all excited states serving as a basis for the subsequent dynamics must be computed
at the same level of theory and using the same atomic basis.
The toolbox \textsc{detCI@ORBKIT} is written in Python, which offers a broad set of efficient standard libraries
and simplifies its portability on different platforms.
The structure of the program can be summarized as follows:
\begin{enumerate}
 \item A parser routine extracts the information about the molecular geometry, the atomic basis, the coefficients of the molecular orbitals and their energies,
          the MO occupation patterns for all desired electronic states, and the $N$-electron eigenstate energies. Currently, the program supports the
          MOLPRO, PSI4, GAMESS, and TURBOMOLE formats. A full and updated list can be found in the program documentation.
 \item The molecular orbitals and the derivatives thereof are reconstructed from the atomic orbitals and projected on an arbitrary grid
          using the functionalities of \textsc{ORBKIT}. All integrals required for computing the matrix elements of the one-electron operators
          in the molecular orbital basis are computed analytically via the underlying atomic basis. These are stored in a Python list for later use.
 \item A library of transition moments is built for each pair of $N$-electron stationary wave functions used in the time-dependent wave packet expansion (cf. Eq.~\eqref{general_CI}).
          The evaluation of transition moments of one-electron operator between two multi-determinant states,
         $\Braket{\Phi_{\lambda}^{{\rm CI}} \left| \hat{F} \right| \Phi_{\nu}^{{\rm CI}}}$, is performed in three steps:
         \begin{enumerate}
           \item The Slater determinants are compared to each other to determine all matrix elements $\Braket{\phi_{p} \left| \hat{F} \right| \phi_{q}}$ involved.
                    The configurations, where both CI-coefficients $\left\{C_{p}^{\left(\lambda\right)},C_{q}^{\left(\nu\right)}\right\}$
                    are larger than a user-defined threshold, are brought to maximum coincidence.
                    The necessary number of orbital permutations determines the phase factor for the rearrangement.
                    Since the number of Slater-determinants in a stationary state can become prohibitively high, the comparison and ordering routine is implemented
                    in Cython\cite{cython} and can be executed on multiple processors.
           \item From the occupation pattern two cases are identified: two identical determinants, and two determinants that differ by a single spin orbital.
                    These build an integer list of important contributions.
                    All one-electron integrals between two spin orbitals, $\Braket{ \varphi_a \left| \hat{F}\right| \varphi_a}$ and
                    $\Braket{ \varphi_a \left| \hat{F}\right| \varphi_r}$, which are necessary to calculate the
                    expectation value of the operator, are loaded from the MO Python list generated by \textsc{ORBKIT} in step 2.
           \item The transition moments $\Braket{\Phi_{\lambda}^{{\rm CI}} \left| \hat{F} \right| \Phi_{\nu}^{{\rm CI}}}$
                    in Eq.~\eqref{exp_val_2} are calculated by weighting the integrals over the spin orbitals with the associated CI-coefficients
                    $\left\{C_{p}^{\left(\lambda\right)},C_{q}^{\left(\nu\right)}\right\}$.
                    In general, weighting the CI-coefficients and the phase factor with the grid-dependent molecular orbitals is the bottleneck of
                    our routine, and significant effort was made on the optimization and parallelization of the associated Cython\cite{cython} modules.
         \end{enumerate}
 \item The library can then be kept in memory or stored on disk for later use. The Python architecture allows direct visualization of the one-electron quantities
          using packages such as matplotlib\cite{matplotlib} or Mayavi\cite{mayavi}.
          In addition, various file formats (e.g., standard Gaussian cube-files, hierarchal HDF5 data files, etc.) are supported,
          which enables visualization with external graphical programs, such as, VMD\cite{vmd} or ZIBAmira\cite{Amira}.
\end{enumerate}
The elements in the CI-determinant basis yield a matrix representation of the electronic Hamiltonian, which can be used to investigate
the $N$-electron dynamics of a wave packet of the form Eq.~\eqref{td_wf}. The analysis is performed by weighting with the time-dependent
coefficients $B_{\lambda}{\left(t\right)}$ the expectation values and transition moments involved in a desired operator. These are
the position representation for the one-electron density, Eq.~\eqref{rho}, 
and the transition moments, Eq.~\eqref{transitionJ}, for the electronic flux density, Eq.~\eqref{j}.
The dynamical program is not part of the standard \textsc{detCI@ORBKIT} implementation.

\section{Numerical Examples}\label{examples}

In this section, the capabilities of our toolbox
to study the correlated electron dynamics in real-time are illustrated for different molecular systems.
Two test molecules are studied to reveal the dependence of the computational procedure towards 
the quality of the electronic structure theory and influence of the basis set: the trihydrogen 
cation H$_{3}^{+}$ and the lithium hydride molecule, LiH.
A Python execution code for each of these examples and the data from the associated quantum chemical calculation 
are available in the program package.
Note that a development version of \textsc{detCI@ORBKIT} was already used for several applications in the 
literature, see Refs.~\cite{15:HT:gesi,hermann2016alianatase,dongming2016benzene}.

\subsection{Basis Set Dependence of the Continuity Equation}\label{h3+}

We advocate using the time-dependent electron density
$\rho\left(\mathbf{r},t\right)$ and time-dependent electronic flux density $\mathbf{j} \left(\mathbf{r},t\right)$
as complementary quantities for the analysis of correlated electron dynamics in molecular systems.
The dependence of an $N$-electron dynamics on the underlying atomic basis set is an important 
convergence parameter that determines the quality of this analysis.
To assess the robustness of the predictions concerning the time-dependent electronic flux density towards the
basis set size, the trihydrogen cation H$_{3}^{+}$ is studied using the minimal basis set, 
STO-3G, as well as a systematic series of Dunning-type basis sets, cc-pVXZ and aug-cc-pVXZ with X=D, T, Q, 5.\cite{hehre1969program,dunning1989ccpvxz}
In these calculations, the H$_{3}^{+}$ molecule is chosen to retain the equilateral triangular equilibrium geometry ($r_{\rm HH}=0.87\, {\rm \AA}$)
of the ground state $1\,{}^1A_1'$ aligned in the $xy$-plane.
To avoid artifacts coming from the electronic structure method, 
all calculations are performed at the Full CI level, using the open-source package PSI4\cite{psi4}.

The reasons for selecting H$_{3}^{+}$ as a test system is threefold:
First, the quantum chemical calculations can be performed at the Full CI limit for a large 
selection of basis sets due to the small number of electrons in H$_{3}^{+}$.
Second, its electronic structure is already well-studied due to its important 
role in interstellar chemistry.\cite{geballe1996detection,oka2006interstellar,larsson2008h} 
This enables to compare with high-quality reference calculations.\cite{Pavanello2009H3+}
Finally, initial conditions can be chosen such as to drive an interesting charge migration process
that induces a periodic unidirectional circular current in the H$_{3}^{+}$ plane.
As was previously shown for other high-symmetry ring-shaped molecules \cite{barth2006periodic,dongming2016benzene,Hermann2017},
this can be achieved by a carefully chosen superposition of the ground state $1\,{}^1A_1'$ and the degenerate state $1\,{}^1E'$.
In the basis of Full CI eigenstates used to study the $N$-electron dynamics, the field-free evolution of the system is 
known analytically at all times.
In the particular case presented here, the time evolution of a wave packet consisting of two superposition states takes the following form
\begin{eqnarray}\label{wf_2st}
  \Ket{\Psi_{{\rm el}}\left(t\right)} = \frac{1}{\sqrt{2}} \left(\Ket{\Psi_{\rm g}} e^{-i  E_{\rm g} t/\hbar} +\Ket{\Psi_{\rm e}} e^{-i (E_{\rm e} t/\hbar + \eta)}\right),
\end{eqnarray}
where $\Ket{\Psi_{\rm g}}$ is the stationary wave function of the ground 
state $1\,{}^1A'$, and $\Ket{\Psi_{\rm e}}$ denotes the stationary wave function of the degenerate excited state $1\,{}^1E'$.
The latter is chosen as a complex-valued linear combination of $1\,{}^1E_x'$ and $1\,{}^1E_y'$ with the relative phase set to $\eta=0$
\begin{eqnarray}
  \Ket{\Psi_{\rm e}} = \frac{1}{\sqrt{2}} \left(\Ket{\Psi_x} +i  \Ket{\Psi_y}\right).
\end{eqnarray}
Here, $\Ket{\Psi_x}$ and $\Ket{\Psi_y}$ refer to the wave functions of state $1\,{}^1E_x'$ and $1\,{}^1E_y'$, respectively.
The axis labels correspond to the molecular orientation as given in the quantum chemistry program.
It was shown that it is possible to prepare such a wave packet by electronic excitation of the ground state using a circular polarized laser field.\cite{dongming2016benzene}
The time-dependent electron density associated with this wave packet takes the form
\begin{eqnarray}\begin{aligned}\label{rho_h3+}
  \rho\left(\mathbf{r},t\right) =& \frac{1}{2} \Big(\rho_{\rm g}\left(\mathbf{r}\right) +\frac{1}{2}\rho_{x}\left(\mathbf{r}\right)+\frac{1}{2}\rho_{y}\left(\mathbf{r}\right) \Big)
  + \frac{1}{\sqrt{2}}\rho_{{\rm g}x}\left(\mathbf{r}\right)\cos(\Delta E t/\hbar)\\
  &+ \frac{1}{\sqrt{2}}\rho_{{\rm g} y}\left(\mathbf{r}\right)\sin(\Delta E t/\hbar),
\end{aligned}\end{eqnarray}
with $\Delta E = E_{\rm{e}}-E_{\rm{g}}$.
The first term on the right-hand-side describes a static contribution to the one-electron density, where
$\left\lbrace\rho_{\rm g},\rho_{x},\rho_{y} \right\rbrace$ are the respective contributions from the ground state $1\,{}^1A_1'$
and the excited states $1\,{}^1E_x'$ or $1\,{}^1E_y'$. 
The evolution of the electronic wave packet is driven by the  transition electron densities, $\left\lbrace \rho_{{\rm g}x},\rho_{{\rm g} y} \right\rbrace$,
which are obtained by resolving Eq.~\eqref{rho} in the basis of Full CI eigenstates.
Similarly, the time-dependent electronic flux density obtained from Eq.~\eqref{j} reads
\begin{eqnarray}\begin{aligned}\label{j_h3+}
  \mathbf{j} \left(\mathbf{r},t\right) =& -\frac{1}{2} \mbox{Im}\left[\mathbf{j}_{xy} \left(\mathbf{r}\right)\right]
  + \frac{1}{\sqrt{2}} \mbox{Im}\left[\mathbf{j}_{\rm{g}x} \left(\mathbf{r}\right)\right] \sin(\Delta E t/\hbar)
  - \frac{1}{\sqrt{2}} \mbox{Im}\left[\mathbf{j}_{gy} \left(\mathbf{r}\right)\right] \cos(\Delta E t/\hbar),
\end{aligned}\end{eqnarray}
where $\left\lbrace \mathbf{j}_{{\rm g}x},\mathbf{j}_{{\rm g}y}, \mathbf{j}_{xy} \right\rbrace$ 
stand for the transition electronic flux densities between the ground and the excited states.
For a longer derivation of Eqs.~\eqref{rho_h3+},\,\eqref{j_h3+}, the reader is referred to previous work. \cite{dongming2016benzene}

\begin{figure}
\centering
\includegraphics[width=14cm]{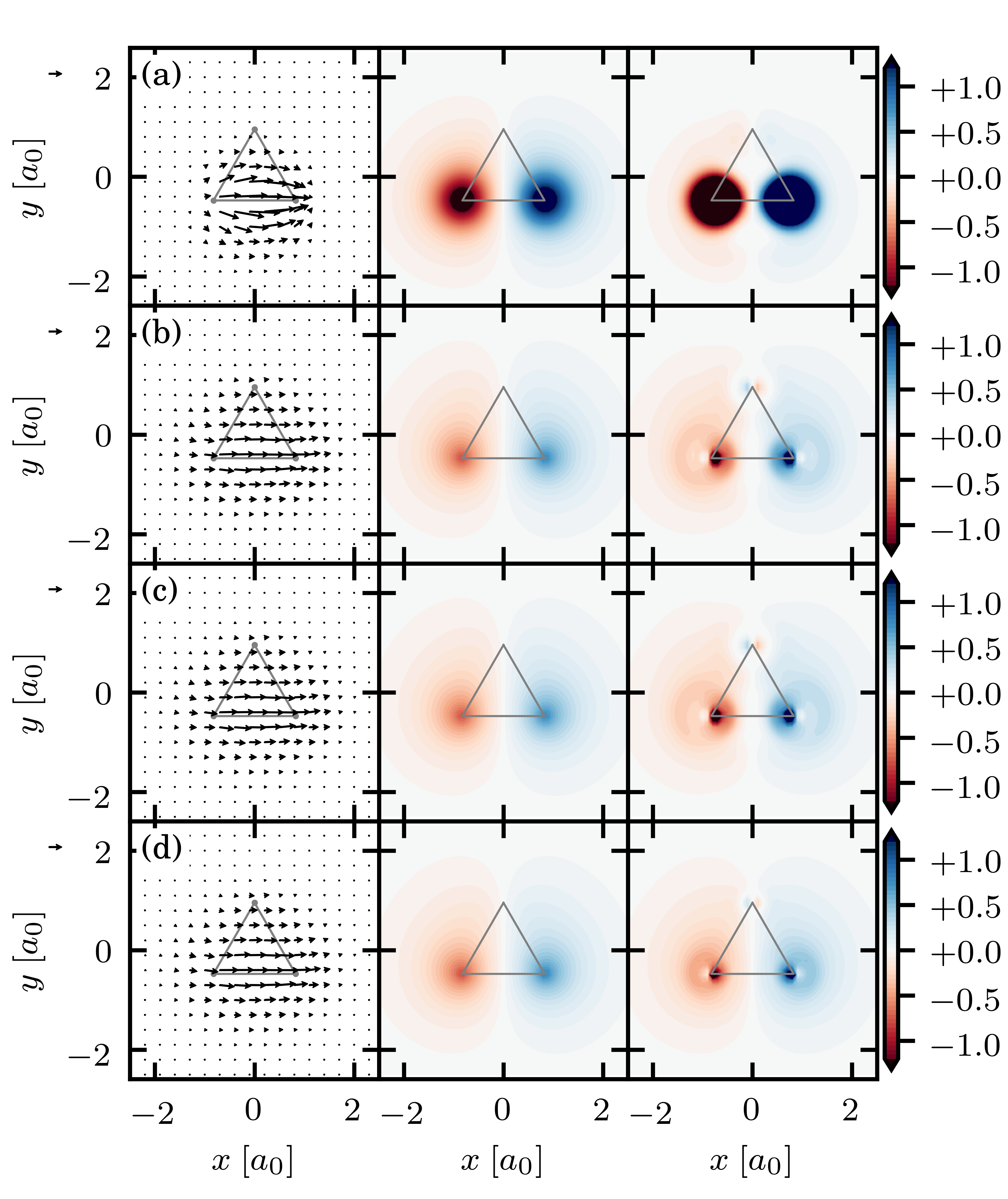}
\caption{
Vector plots of the electronic flux density $\mathbf{j} \left(\mathbf{r},t\right)$ (left panels)
and contour plots of the time derivative of the electron density $\partial \mathbf{\rho} \left(\mathbf{r},t\right) / \partial t$ 
(central panels) and of the divergence of the flux density $- \vec{\nabla} \cdot \mathbf{j} \left(\mathbf{r},t\right)$ 
(right panels) (cf. Eq.~\eqref{con_eq}) for the trihydrogen cation H$_{3}^{+}$ at $t=\tau/4$
based on Full CI calculations. 
A comparison between different basis sets is illustrated: (a) STO-3G, (b) cc-pVTZ, 
(c) aug-cc-pVTZ, and (d) aug-cc-pV5Z.
The trihydrogen cation H$_{3}^{+}$ is represented as a gray stick model.
A reference arrow with a length of $5\cdot 10^{-2}E_{\rm h } / \hbar a_{0}^{2}$ is shown at the ordinate.
The contour plots of the time derivative of the electron density $\partial \mathbf{\rho} \left(\mathbf{r},t\right) / \partial t$ 
and of the divergence of the flux density $- \vec{\nabla} \cdot \mathbf{j} \left(\mathbf{r},t\right)$ 
are in units of $10^{-1} E_{\rm h} / \hbar a_{0}^{3}$. }
\label{h3+_ceq}
\end{figure}

In order to determine the dependency of the electron density and the transition flux density towards the quality of the basis set, 
we make use of the continuity equation.
The time derivative of the electron density $\partial \mathbf{\rho} \left(\mathbf{r},t\right) / \partial t$
(left-hand side of Eq.~\eqref{con_eq}) will be hereafter referred to as ``electron flow'',
to differentiate this quantity from the divergence of the 
electronic flux density $- \vec{\nabla} \cdot \mathbf{j} \left(\mathbf{r},t\right)$ (right-hand side of Eq.~\eqref{con_eq}).
Both quantities represent expressions for the electron flow and should converge to the same value by definition.
Since Full CI calculations are exact for a given basis set, any difference between both solely stems from the quality of the basis set.
For clarity, only a few of the basis sets mentioned above are compared here.
These include the minimal basis set STO-3G, and three correlation consistent basis sets, cc-pVTZ, aug-cc-pVTZ, and aug-cc-pV5Z.
The results of 
Fig.~\ref{h3+_ceq} show the electron flow $\partial \mathbf{\rho} \left(\mathbf{r},t\right) / \partial t$ (central panels),
the transient electronic flux density $\mathbf{j} \left(\mathbf{r},t\right)$ (left panels),
and its divergence $- \vec{\nabla} \cdot \mathbf{j} \left(\mathbf{r},t\right)$ (right panels) 
for the superposition of the ground state $1\,{}^1A_1'$ and the excited state $1\,{}^1E'$ at $t=\tau/4$ ($\tau = h/\left( \Delta E \right)$).
The red (blue) areas in the central and right panels represent regions of instantaneous decrease (increase) of the electron density.
At first glance, it can be observed that the qualitative features of the electron 
flux density and the electron flow are very robust towards the basis set quality.
That is, the electrons migrate from the lower left hydrogen atom to the lower right one.
Further, the s-character of the three orbitals involved in the H$_3^+$ bond
can be quite easily recognized. This character is retained when using more complete basis sets for the analysis of the electron flow based on the density.
As a complementary analysis tool, the electronic flux density maps reveal that the density migrates from one hydrogen to the other
along along the bond at this time step. This somewhat counterintuitive feature is recognized
for all basis sizes.

Nonetheless, several differences between $\partial \mathbf{\rho} \left(\mathbf{r},t\right) / \partial t$ 
and $- \vec{\nabla} \cdot \mathbf{j} \left(\mathbf{r},t\right)$ can be identified across the different basis sets.
On the one hand, the electron flow and the vorticity in the electronic flux density obtained with the minimal basis set
(STO-3G, top panels) are much larger than the results from the Dunning-type basis sets.
This is due to the small number of basis functions, which overestimates the contribution of the $s$-orbitals to the
total density. The electron flow appears to be converged already at the cc-pVTZ level (second line). 
When comparing $\partial \mathbf{\rho} \left(\mathbf{r},t\right) / \partial t$ and $- \vec{\nabla} \cdot \mathbf{j} \left(\mathbf{r},t\right)$,
artifacts can be observed around the hydrogen nuclei.
Since the electron density $\mathbf{\rho} \left(\mathbf{r},t\right)$ is rather insensitive 
towards the basis set quality, these artifacts merely occur when computing the divergence of the electronic flux density,
$- \vec{\nabla} \cdot \mathbf{j} \left(\mathbf{r},t\right)$.
Using the cc-pVDZ basis (not shown), the nodal structures at the nuclei are largest and they disappear slowly as the number of basis functions
increases. The reason for the slow convergence of the  divergence of the electronic flux density in the present example is that it is dominated by 
the derivative of the density at the nuclei. Since the cusp at a nucleus is poorly represented using Gaussian-type atomic orbitals, 
the derivative of the transient flux density is mostly affected.

\begin{figure*}
\centering
\includegraphics[width=16cm]{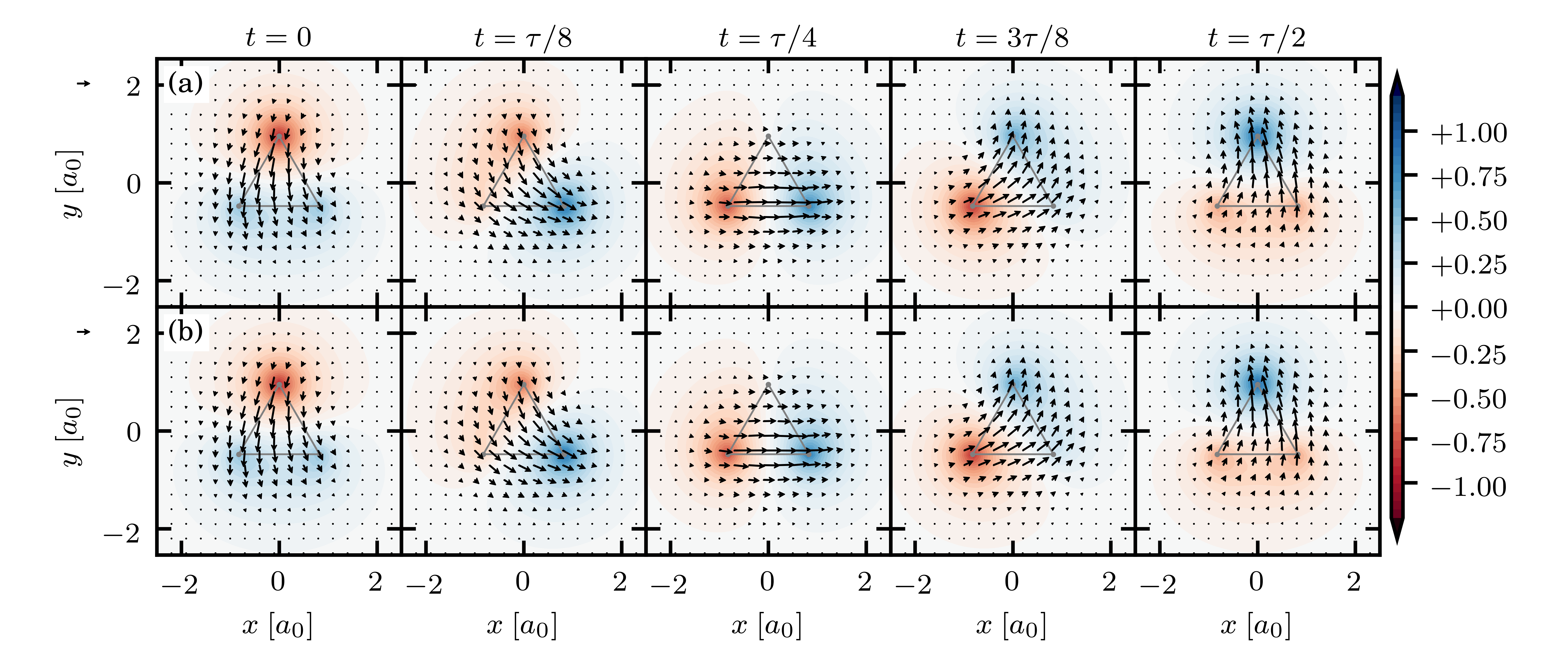}
\caption{Representative snapshots of the electronic flux density $\mathbf{j} \left(\mathbf{r},t\right)$ 
(in units of $10^{-1} E_{\rm h} / \hbar a_{0}^{2}$) and of the electron flow 
$\partial \mathbf{\rho} \left(\mathbf{r},t\right) / \partial t$ 
(in units of $E_{\rm h} / \hbar a_{0}^{3}$) for the trihydrogen cation 
H$_{3}^{+}$ at different times on the basis of Full CI calculations. 
The results are plotted for two different basis sets:
(a) cc-pVTZ and (b) aug-cc-pV5Z.
The structure of the trihydrogen cation H$_{3}^{+}$ is presented as a gray 
ball-and-stick model.
A reference arrow with a length of $5\cdot 10^{-2}E_{\rm h } / \hbar a_{0}^{2}$ is shown at the ordinate.}
\label{h3+_tdfd}
\end{figure*}

The same phenomenological robustness is observed for the time-evolution of the electronic flux density
$\mathbf{j} \left(\mathbf{r},t\right)$ and the electron flow, $\partial \mathbf{\rho} \left(\mathbf{r},t\right) / \partial t$.
In Fig.~\ref{h3+_tdfd}, the time-dependent flux densities (arrows) are superimposed on the time-dependent electron flow,
where the red (blue) areas indicate regions of decreasing (increasing) density.
They are depicted at characteristic times during half the period $\tau$ of the charge migration process, $0 \leq t \leq \tau/2$. 
The period $\tau$ is related to the energy difference $\Delta E$ as follows $\tau = h/\Delta E$.
The results are shown for the cc-pVTZ (top panels) and the aug-cc-pV5Z basis set (bottom panels),
for which the energy difference is $\Delta E=19.36\,{\rm eV}$ and $\Delta E=19.33\,{\rm eV}$, respectively.
Both tools correctly predict a clockwise circular migration of the electron 
for the transition between state $1\,{}^1A_1'$ and state $1\,{}^1E'$.
Neither qualitative nor quantitative differences between the two basis sets can be recognized.
From Eq.~\eqref{j_h3+}, it can be observed that the electronic flux density has three components:
a static ring current $\mathbf{j}_{xy}$, and two alternating polarized components $\mathbf{j}_{gx}$ and $\mathbf{j}_{gy}$.
The latter two create the asymmetric pattern observed in the flux density, which shows that the density
hops from atom to atom along an inward curved path (see, e.g. at $t=\tau/8$).
Quite importantly, this information cannot be obtained from neither the density nor the electron flow.

\begin{figure}
\centering
\includegraphics{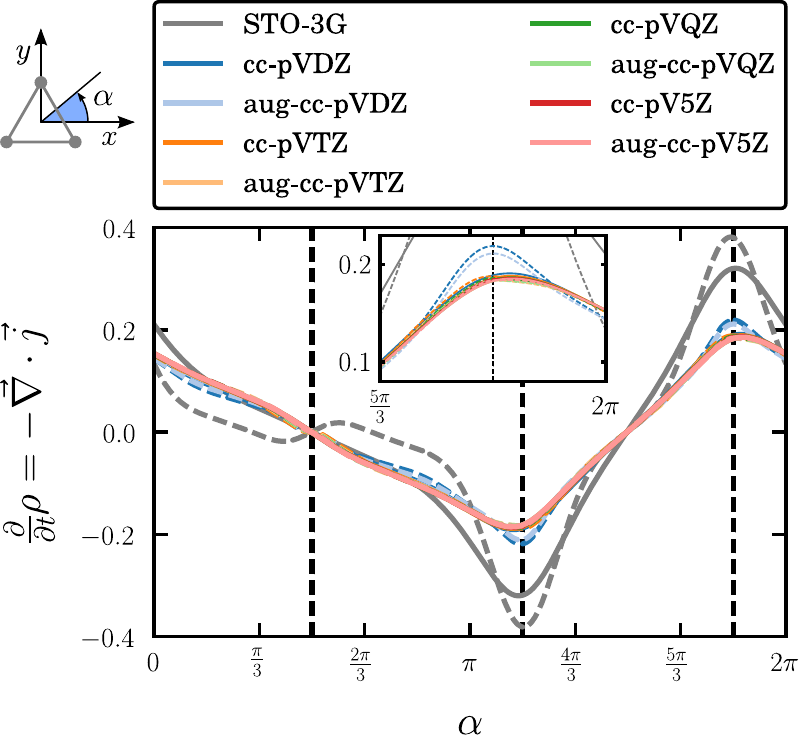}
\caption{Time derivative of the electron density $\partial \mathbf{\rho} \left(\mathbf{r},t\right) / \partial t$ 
(solid lines) and divergence of the flux density $- \vec{\nabla} \cdot\mathbf{j} \left(\mathbf{r},t\right)$ 
(dashed lines) (cf. Eq.~\eqref{con_eq}) as a function of the polar angle $\alpha$ at $t=\tau/4$ for the trihydrogen cation 
H$_{3}^{+}$ for selected basis sets at the Full CI level.
The orientation of the polar angle $\alpha$ with respect to H$_{3}^{+}$ is defined 
in the sketch alongside to the legend.
The inset shows an enlarged view of the graph from $\alpha = 5/3\pi$ to $\alpha = 2\pi$.
The three vertical dashed black lines mark the angular positions of the hydrogen
nuclei.
The plotted quantities, i.e., $\partial \mathbf{\rho} \left(\mathbf{r},t\right) / \partial t$ 
and $- \vec{\nabla}\cdot \mathbf{j} \left(\mathbf{r},t\right)$,
are in units of $E_{\rm h} / (\hbar \,{\rm rad})$.}
\label{h3+_ceq_1d}
\end{figure}

In order to accurately determine the influence of the basis set size on the electron flow
($\partial \mathbf{\rho} \left(\mathbf{r},t\right) / \partial t$, solid lines)
and the divergence of the electronic flux density ($- \vec{\nabla} \cdot \mathbf{j} \left(\mathbf{r},t\right)$, dashed lines),
both quantities are illustrated in Fig.~\ref{h3+_ceq_1d} as a function of the polar angle $\alpha$ at $t=\tau/4$.
The functions are obtained by projecting the electronic density on a cylindrical grid, $\{r,\alpha,z\}$, and integrating 
both sides of the continuity equation over $r\,{\rm d}r\,{\rm d}z$. The grid representation is here again produced using \textsc{ORBKIT}
and the integrals are performed via cubature \cite{pycubature,pub1_cubature,pub2_cubature}. The vertical dashed black lines denote the position of the nuclei.
As previously observed, the minimal basis STO-3G is seen to poorly satisfy the continuity equation, as the discrepancies
between the two quantities (gray curves) remain large over the whole domain $[0,2\pi]$.
In comparison, the Dunning-type basis sets perform very well over the whole angular range already at the cc-pVDZ level.
A look at the angular electron flow in the vicinity of the nuclei  (cf. inset in Fig.~\ref{h3+_ceq_1d}) reveals that
the plotted quantities ($\partial \mathbf{\rho} \left(\mathbf{r},t\right) / \partial t$ and $- \vec{\nabla} \cdot \mathbf{j} \left(\mathbf{r},t\right)$)
-- and with them the electronic continuity equation Eq.~\eqref{con_eq} -- 
are quantitatively converged from the cc-pVTZ basis set level.

\begin{figure}
\centering
\includegraphics[width=0.6\textwidth]{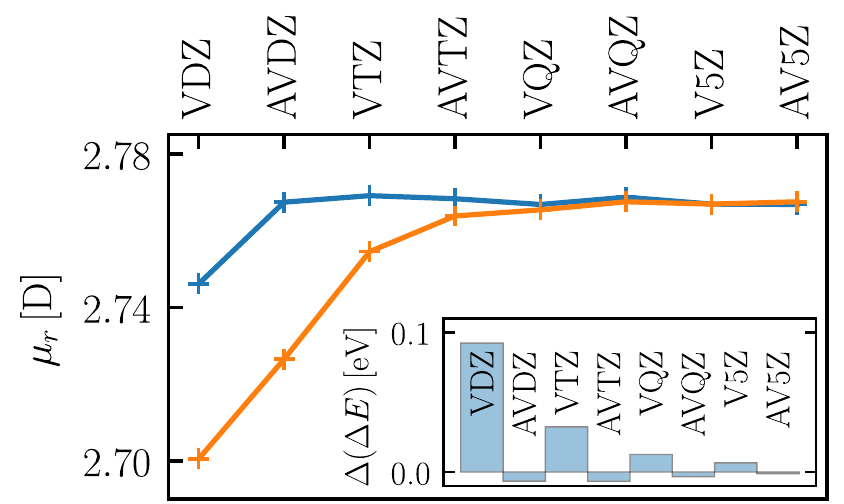}
\caption{Comparison between transition dipole moments in length gauge $\mu_r$ and 
transition dipole moments in length gauge calculated from the dipole moment in 
velocity gauge $(\mu_v)_r$ (cf. Eq.~\eqref{mu_vg_to_lg}) for the transition between 
state $1 {}^1A_1'$ and state $1 {}^1E'$.
The dipole moments from Full CI calculations are depicted for different Dunning basis sets.
These are sorted with increasing number of basis functions, (i.e., VXZ stands for cc-pVXZ, and
AVXZ symbolizes aug-cc-pVXZ)
The inset shows the difference between the calculated excitation energy
for the selected basis sets and a reference excitation energy ($\Delta E = 19.3289$ eV).}
\label{fig_dm_h3+}
\end{figure}

A further quantitative measure for the basis set convergence can be obtained by comparing the dipole moment in length gauge (cf. Eq.~\eqref{mu_op_lg}) 
with the corresponding one calculated from the dipole moment in velocity gauge (cf. Eq.~\eqref{mu_vg_to_lg}).
Both quantities must be equal for a converged basis set in a Full CI calculation.
In Tab.~\ref{h3+_dm}, the $x$-components of the transition dipole moment for the two gauges are listed,
along with the associated excitation energies for the $\Ket{\Psi_g}\to\Ket{\Psi_x}$ state transition
and the total number of atomic basis functions $N_{\rm AO}$ for each basis set. 
The number of Full CI configurations for each state is obtained from the determinental CI output of the quantum chemistry program.
This number is smaller for the ground state because all single excitations are projected out.
The degenerate excited state exhibits a small splitting due to the use of the D$_{2h}$ abelian symmetry group in the calculation.
Apart from the poor STO-3G basis, all results obtained from correlation consistent basis sets are quite homogeneous and yield
a smooth convergence towards the literature benchmark value. As was the case for the electronic flux density, the \hbox{cc-pVTZ}
values appear to be converged to a sufficient accuracy to allow for quantitative analysis. A similar level of accuracy on the
energy and the dipole moment is obtained with a marginally smaller basis using diffuse functions, at the aug-cc-pVDZ level.
Note that increasing the basis size also increases the number of Full CI configurations (see last columns of Tab.~\ref{h3+_dm}),
which improves the variational description of the molecular orbitals and the $N$-electron wave functions simultaneously.
These observations are complemented by Fig.~\ref{fig_dm_h3+}, which shows the dipole moments and excitation 
energies as a function of the basis set size, ordered according to the number of functions.
Here, the excitation energies are plotted as the difference energy to a highly accurate reference energy.
The minimal basis set is excluded for clarity.
From the inset in Fig.~\ref{fig_dm_h3+}, it can be deduced that diffuse functions 
play a more important role for the energy convergence than adding basis functions with higher angular 
momenta.
Interestingly, this does not apply to the convergence of the two dipole moments in length gauge,
Eqs.~\eqref{mu_op_lg} and \eqref{mu_vg_to_lg},
which converge monotonically with the number of basis functions.

\subsection{Impact of the Electronic Structure Method}\label{lih}

As a second example, the impact of different electronic structure methods on the time-dependent electron density and electronic flux density is investigated.
In particular, both quantities are computed on the basis of a Full CI calculation and compared with those obtained from a CI Singles calculation, from a restricted 
active space configuration interaction (RASCI), and from Complete-Active Space Self-Consistent Field (CASSCF) calculations.
The details of the active space are given below.
These methods build a hierarchy of electronic structure theories, where the description of electron correlation is improved systematically
by considering either a larger active space or a higher degree of excitations.
Since a Full CI calculation is used as a reference, a well-studied four-electron molecule, the heteronuclear polar lithium hydride LiH, is 
chosen as a test system.\cite{Docken1972,Stwalley1993,Gadea2006,remacle2007laser,nest2008pump,Bande2010347,Mignolet6721,nikodem2016quantum}
We are particularly interested in the charge transfer state A${}^1 \Sigma^+$, which is optically accessible from the electronic ground state
X${}^1 \Sigma^+$ and lies in the Franck-Condon region. 
By using a so-called $\pi/2$-pulse, it is possible to create a superposition state as in Eq.~\eqref{wf_2st},
where $\Ket{\Psi_{\rm g}}$ is the stationary wave function of the ground state and $\Ket{\Psi_{\rm e}}$ that of the charge transfer state.

The  one-electron density associated with this superposition state evolves in time according to 
\begin{eqnarray}\begin{aligned}\label{rho_lih}
  \rho\left(\mathbf{r},t\right) =& \frac{1}{2} \Big(\rho_{\rm g}\left(\mathbf{r}\right) + \rho_{\rm e}\left(\mathbf{r}\right)\Big)
  + \rho_{\rm ge}\left(\mathbf{r}\right)\cos(\Delta E\,t/\hbar + \eta),
\end{aligned}\end{eqnarray}
where $\Delta E=E_{{\rm e}} -E_{{\rm g}}$. The static one-electron density of the ground state X${}^1 \Sigma^+$ ($\rho_{\rm g}\left(\mathbf{r}\right)$)
and charge transfer state A${}^1 \Sigma^+$ ($\rho_{\rm e}\left(\mathbf{r}\right)$), as well as the static transition density ($\rho_{\rm ge}\left(\mathbf{r}\right)$),
are computed using Eq.~\eqref{rho_CI_mo}.
Similarly, the electronic flux density takes the simplified form
\begin{eqnarray}\begin{aligned}\label{j_lih}
  \mathbf{j} \left(\mathbf{r},t\right) =  \mbox{Im}\left[\mathbf{j}_{\rm ge} \left(\mathbf{r}\right)\right] \sin(\Delta E\, t/\hbar + \eta),
\end{aligned}\end{eqnarray}
where the transition electronic flux density $\mathbf{j}_{\rm ge} \left(\mathbf{r}\right)$ is computed using Eq.~\eqref{j_CI_mo}.
Note that, contrary to the previous example, the electronic flux density does not have a time-independent current term (cf. Eq.~\eqref{j_h3+}).
To complement the analysis of electron migration, the difference density\cite{barth2009concerted,Berg} is calculated
for the specific superposition state from Eqs.~\eqref{yield}, \eqref{rho_lih}, and the choice of phase $\eta=\pi$ as
\begin{eqnarray}\begin{aligned}\label{yield_lih}
  \mathbf{y} \left(\mathbf{r},t \right) &= \rho \left(\mathbf{r},t \right) - \rho \left(\mathbf{r},t=0 \right) \\
  &= \left(1- \cos(\Delta E\,t/\hbar) \right) \rho_{\rm ge}\left(\mathbf{r}\right).
\end{aligned}\end{eqnarray}
As the electron flow itself, it is found to be independent of the static densities of the ground and excited states.

\begin{figure*}
\centering
\includegraphics{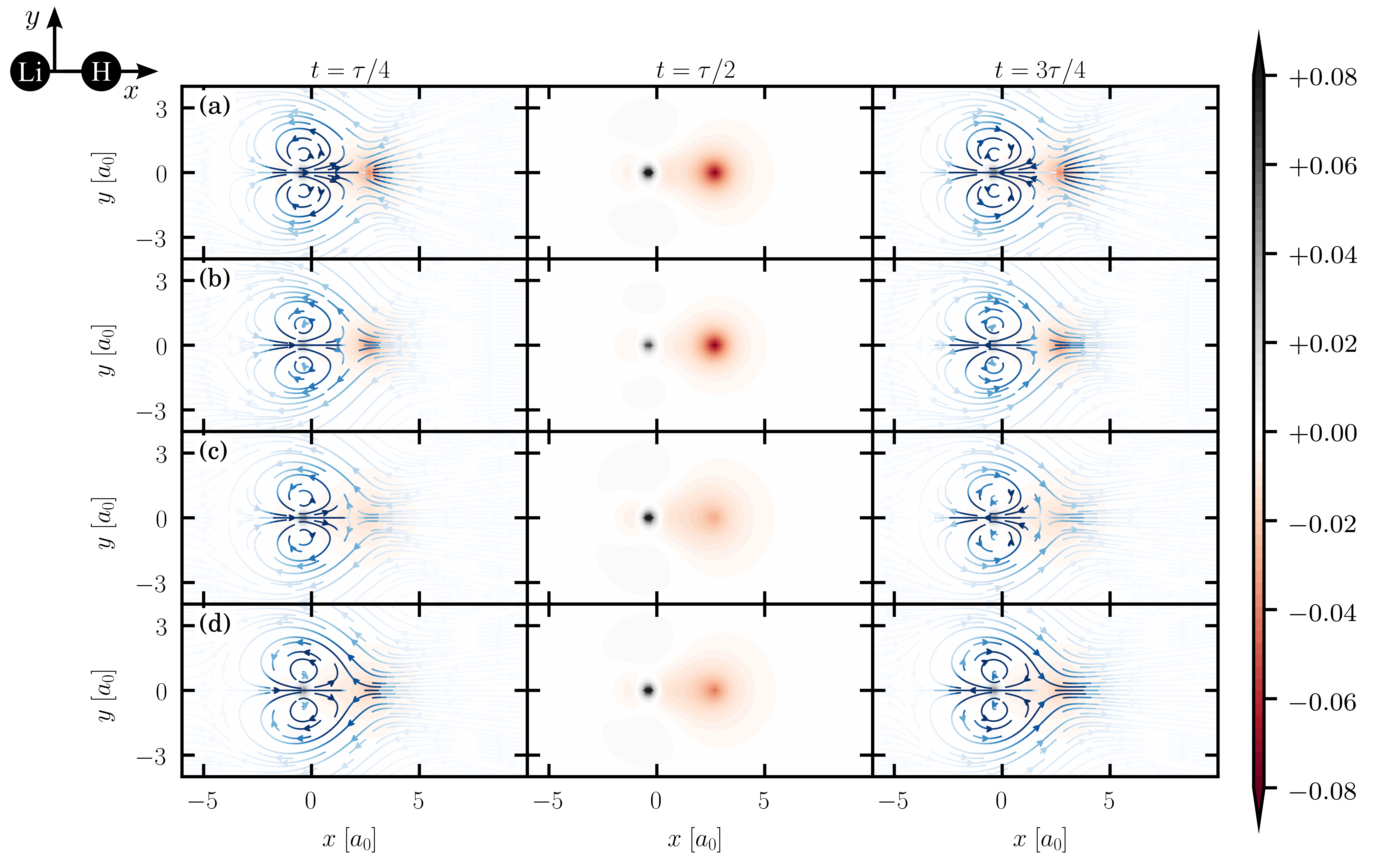}
\caption{Representative snapshots of the flux density $\mathbf{j} \left(\mathbf{r},t\right)$ 
and the difference density $\mathbf{y} \left(\mathbf{r},t\right)$ for the lithium hydride molecule
LiH oriented along the $x$-axis.
Single-point calculations for different levels of electronic structure theory are 
compared including
(a) a CI Singles calculation,
(b) a RASCI(2,5) calculation,
(c) a state-averaged CASSCF(2,5) calculation, and
a Full CI calculation.
The results are obtained using an aug-pVTZ basis set for the superposition of the 
ground state X${}^1 \Sigma^+$ and the first excited state A${}^1 \Sigma^+$.
The flux densities $\mathbf{j} \left(\mathbf{r},t\right)$ are in units of 
$E_{\rm h} / \hbar a_{0}^{2}$, and the difference densities $\mathbf{y} \left(\mathbf{r},t\right)$
are in units of $1 / a_{0}^{3}$.
}
\label{lih_tdfd_y}
\end{figure*}

For each electronic structure method, a single-point calculation at the ground-state equilibrium 
geometry of LiH ($r_{\rm Li-H}=1.63\,{\rm \AA}$) are performed with PSI4\cite{psi4} using an aug-cc-pVTZ basis set.
On this basis, the time-dependent electronic flux densities and density differences
are calculated at representative time steps during one period $\tau$ ($\tau = h/\left( \Delta E \right)$) 
of the charge migration: $\tau/4$, $\tau/2$, and $3\tau/4$.
This choice allows for a direct comparison of the dynamics despite the different transition
energies, and hence the different timescales, found using the various methods.
The resulting $\mathbf{j} \left(\mathbf{r},t \right)$ and $\mathbf{y} \left(\mathbf{r},t \right)$ 
are depicted in Fig.~\ref{lih_tdfd_y}.
The four horizontal panels show the results for the different levels of electronic 
structure methods: (a) CIS, (b) RASCI(2,5), (c) CASSCF(2,5), and (d) Full CI.
The energy difference between the ground and charge transfer states are found to be 
$\Delta E_{\rm CIS}=4.04\,{\rm eV}$, $\Delta E_{\rm RAS(2,5)}=4.13\,{\rm eV}$,
$\Delta E_{\rm CAS(2,5)}=3.41\,{\rm eV}$, and $\Delta E_{\rm FCI}=3.56\,{\rm eV}$.
By definition, $\mathbf{j} \left(\mathbf{r},t \right)$ and $\mathbf{y} \left(\mathbf{r},t \right)$ 
are zero at $t=0$ and $t=\tau$ and therefore, not depicted here.

In the example presented, the Full CI method serves as a reference, since it provides 
the exact solution for the correlated electronic wave function within the limitation 
of a finite basis set. We will thus first proceed to a qualitative analysis of the associated
electronic flux density and density difference to highlight their main characteristics.
The regions of electron density depletion are denoted in red and the electronic flux density is depicted
using streamlines, an alternate representation that connects the arrows of the vector field to 
give the impression of a fluid in motion.
Phenomenologically, an electron migration from the hydrogen atom to the lithium atom is observed
during the first half period $\tau/2$, and the reverse process for the second half (cf. Fig.~\ref{lih_tdfd_y}).
This qualitative observation is indeed expected for the superposition of the more ionic ground
state X${}^1 \Sigma^+$ with the first excited state A${}^1 \Sigma^+$ which has a more covalent character.
As can be seen from the density difference, the electron depletion region surrounding the hydrogen atom
(red areas) are much more diffuse than the area of charge concentration (gray areas) on the lithium.
Contrary to the electron flow (not shown), the density difference retains the same sign in this superposition
state at all times. This implies that an electron migrates to the lithium and back, leaving a hole on the
hydrogen atom and filling it subsequently.
In the earlier stages of the propagation (left panels of Fig.~\ref{lih_tdfd_y}),
the electronic flux density exhibits a large vorticity around the lithium atom.
The charge is transferred indirectly from the hydrogen to the lithium, where the electron density flow
forms a torus and the density is enriched from behind. 
This finding is in agreement with previous theoretical studies on similar molecules \cite{nest2008pump,remacle2007laser,nikodem2016quantum}.
The electronic flux density offers the main advantage of revealing all mechanistic features of the electron
flow at first glance.

The CIS method \cite{foresman1992toward} provides a computationally cheap alternative to Full CI 
that only contains single excitations from the reference wave function (cf. Eq.~\eqref{CI}).
Despite some well-known limitations\cite{dreuw2005single,subotnik2011communication}, 
this simple approach is generally expected to correctly recover the qualitative character
of molecular excited states.\cite{head1995analysis,dreuw2003long}
This can likewise be confirmed by comparing the results for the LiH molecule
(cf. Fig.~\ref{lih_tdfd_y} upper panel) with the Full CI benchmark (cf. Fig.~\ref{lih_tdfd_y} lower panel).
It is striking that all features of both the flux density and of the electronic difference density, can be captured at the CIS level.
In particular, the toroidal structure of the time-dependent electronic flux density is very well captured.
However, quantitative differences are discernible for the density difference (see, e.g., the central panels).
In the present case, the main effect seems to be the localization of the positive charge closer to the hydrogen nucleus.
These discrepancies can be attributed to an insufficient representation of electron correlation at the CIS ansatz.

To investigate this effect further,
alternative determinant-based calculations were performed at the CASSCF level of theory\cite{siegbahn1979generalizations,siegbahn1980comparison,roos1980complete,siegbahn1981complete}.
This special form of the MCSCF method, which is briefly explained in 
section \ref{theory}, incorporates higher-order excitations in the description of the correlated wave function.
While the Full CI and the CIS scheme are very straightforward to set up, the choice 
of the active space for a CASSCF calculation is an art in itself.
To design a first active space, an educated guess can be obtained by considering the molecule in the dissociation limit.
In the ground state at the dissociation limit, the lithium atom is found in the configuration 1s${}^2$2s in the ${}^2S$ state
and the hydrogen atom is in the ${}^2S$ state (1s${}^1$), which yield the following degenerate molecular states: 
a singlet $^1\Sigma^+$ and a triplet $^3\Sigma^+$ state.
These states form the lowest covalent dissociation limit.
The second lowest atomic excitation is the transition of the lithium atom to the 1s${}^2$2p configuration in the ${}^2P^o$ state.
This corresponds to four molecular states: a singlet $^1\Sigma^+$ state, a singlet $^1\Pi$ state, a triplet $^3\Sigma^+$ state,
and a triplet $^3\Pi$ state, which represent the second lowest dissociation limit.
In the dissociation limit, only the 2s and 2p orbitals of the lithium are required to describe difference between the ground state X${}^1 \Sigma^+$
and the first singlet excited state A${}^1 \Sigma^+$.
Consequently, the minimal active space consists of two active electrons in five molecular orbitals with the 1s$^2$ of Li as core orbitals.
Note that this analysis does not strictly apply for the molecule at the ground-state equilibrium geometry, but it serves as an initial guess.

According to the prescription above, the LiH molecule is first calculated using the minimal active space while keeping the orbitals frozen.
This restricted active space configuration interaction ansatz \cite{olsen1988determinant} is labeled according to the CASSCF same notation as RASCI(2,5).
The resulting flux density and the associated electronic difference density are depicted in the second row of Fig.~\ref{lih_tdfd_y}.
Again, the results for the RASCI(2,5) calculation correctly recover the qualitative aspects of the electron redistribution process,
in particular the large vorticity of the toroidal vector field surrounding the lithium atom (see left and right panels).
Nonetheless, the region of density depletion in the electronic difference density is found to be much more localized at the hydrogen nucleus compared to Full CI.
The region of density enrichment to the left of the lithium atom at $t=\tau/2$ has also a smaller spatial extent than in the benchmark.
These are strong indications that electron correlation in LiH is poorly described using RASCI(2,5).

To improve the description of the static correlation, the molecule is recalculated at the state-averaged CASSCF(2,5) level of theory.
In general, state-averaging is required to simultaneously calculate degenerate states of different symmetries on the basis of a single set of optimized molecular orbitals.
In our example, this single set is prerequisite to apply the Slater-Condon rules.
As can be seen from the third row of Fig.~\ref{lih_tdfd_y}, this setup yields both a qualitative and quantitative agreement with the reference 
Full CI results but with a fraction of determinants necessary to describe the correlated wave function.
Provided a proper active space can be constructed for a given molecule, this can amount to very significant computational savings.
Even for this small test system, the Full CI approach comprises 9428 Slater determinants, while
only 11 are required for the CASSCF(2,5) wave function ansatz of the first excited state A${}^1 \Sigma^+$.
This number also compares advantageously to the 80 determinants required at the CIS level.
As a bottom line, it appears that qualitative features of the electron migration process can be obtained at even
relatively crude levels of electronic structure theory (e.g., CIS), but a careful treatment of electron correlation
is required for quantitative predictions. 

\section{Conclusions}\label{concl}

In this paper, we have introduced a general framework for post-processing
determinant-based configuration-interaction wave functions.
The primary goal of this open-source project is to develop a tool for the
characterization and analysis of correlated electron dynamics in molecular
systems, where a wave packet is expanded using static $N$-electron wave functions.
The procedure relies on the numerical determination of transition moments of a set of one-electron operators, 
which yields a time- and space-resolved picture of the $N$-electron dynamics.
These include transition densities, the electronic flux density, and various derived observables.
All quantities required to reconstruct the multi-determinant wave functions are extracted from the output
of standard quantum chemistry packages using Gaussian-type atom-centered basis sets.
The entire procedure is implemented in a novel Python program \textsc{detCI@ORBKIT} 
which extends the functionalities of the post-processing toolbox \textsc{ORBKIT}.
The latter calculates molecular electronic properties from the data of single-determinant
wave functions which is also extracted from quantum chemical calculations.
The new procedure is constructed so that it can principally evaluate transition moments
of any one-electron operator, by taking advantage of Slater-Condon rules to drastically 
reduce the numerical effort.
Emphasis was put on the general applicability, the parallelization of computationally 
demanding steps, and the easy visualization of the results.

In the application examples, we have demonstrated that the selected set of one-electron 
quantities is suitable to characterize the correlated electron dynamics for molecular 
systems in real time.
In particular, analysis of the electron flux density reveals microscopic details about the motion and flow
of the electrons during the investigated dynamical processes at first glance.
Its qualitative analysis has proven very robust towards the choice of electronic structure theory method and 
the quality of the underlying atomic basis set, with the exception of the minimal basis STO-3G.
Comparison of the electron flow (the time derivative of the electron density, $\partial \mathbf{\rho} \left(\mathbf{r},t\right) / \partial t$)
with the divergence of the electronic flux density $- \vec{\nabla} \cdot \mathbf{j} \left(\mathbf{r},t\right)$
reveals a slow convergence of the continuity equation, Eq.~\eqref{con_eq}, with respect to the basis size.
It thus appears preferable to base quantitative predictions on observables derived from the electron density,
in particular on the electron flow and its integral over time, the electronic difference density.
In this respect, it appears that an accurate description of electron correlation is of primal importance.
The tools advocated here appear as complementary for the analysis of $N$-electron dynamics.

\section{Acknowledgment}

The authors gratefully thank the Scientific Computing Services Unit of the 
Zentraleinrichtung f{\"u}r Datenverarbeitung at Freie Universt{\"a}t Berlin for 
allocation of computer time.
The funding of the Deutsche Forschungsgemeinschaft through the Emmy-Noether program 
(project TR1109/2-1) and of the Elsa-Neumann foundation of the Land Berlin are
also acknowledged.
Furthermore, we thank Jhon P{\'e}rez-Torres for fruitful discussions during the 
initiation of this project and Prof. J{\"o}rn Manz for encouraging this paper.

\section{Keywords}

Correlated Electron Dynamics, Slater-Condon Rules, Multi-Determinant 
Wave Function, Electronic Flux Density, Electron Density, Electronic Difference Density,
Electronic Current Density


\bibliographystyle{unsrt}
\bibliography{refs_jct,ci_orbkit_wf}

\newpage
\begin{table}
\centering
\caption{Excitation energies $\Delta E$ and transition dipole moments $\mu_r$ in
length gauge for the trihydrogen cation H$_{3}^{+}$ for the transition between
state $1 {}^1A_1'$ and state $1 {}^1E'$ at the Full CI of theory for a selection of basis
sets.
The dipole moment in length gauge $\mu_r$ is compared to $(\mu_v)_r$, the dipole moment in length gauge 
calculated from the dipole moment in velocity gauge (cf. Eq.~\eqref{mu_vg_to_lg}).
Additionally, the number of basis functions is listed for each basis set.}
\begin{center}
\begin{tabular}{l l l l c c c}
\hline
Basis Set & $\mu_r$ [D] & $(\mu_v)_r$ [D] & $\Delta E$ [eV] & $N_{{\rm AO}}$& $N_{{\rm DET},g}$ & $N_{{\rm DET},e}$\\
\hline
STO-3G 		& 2.9336 & 1.8433 & 23.6135 & \phantom{00}3	& \phantom{000}3	& \phantom{0000000}4/4\\
cc-pVDZ 	& 2.7461 & 2.7005 & 19.4211 & \phantom{0}15	& \phantom{00}45	& \phantom{000000}68/68\\
cc-pVTZ 	& 2.7691 & 2.7545 & 19.3612 & \phantom{0}42	& \phantom{0}300	& \phantom{0000}464/464\\
cc-pVQZ 	& 2.7668 & 2.7654 & 19.3414 & \phantom{0}90	& 1292			& \phantom{00}2011/2000\\
cc-pV5Z 	& 2.7669 & 2.7669 & 19.3354 & 165		& 4161			& \phantom{00}6507/6518\\
aug-cc-pVDZ 	& 2.7674 & 2.7265 & 19.3224 & \phantom{0}27	& \phantom{0}139	& \phantom{0000}212/212\\
aug-cc-pVTZ 	& 2.7683 & 2.7638 & 19.3223 & \phantom{0}69	& \phantom{0}817	& \phantom{00}1248/1222\\
aug-cc-pVQZ 	& 2.7687 & 2.7675 & 19.3256 & 138		& 3046			& \phantom{00}4643/4626\\
aug-cc-pV5Z 	& 2.7668 & 2.7675 & 19.3277 & 240		& 9448			& 13836/13680\\
\hline
Ref. \cite{Pavanello2009H3+} & & & 19.3289  & 600\\
\hline
\end{tabular}
\end{center}
\label{h3+_dm}
\end{table}
\newpage

\clearpage
\newpage
\section{TOC}

\textbf{Unraveling correlated electron dynamics:} We introduce an open-source Python framework
to post-process determinant-based configuration-interaction data from standard quantum chemistry packages.
The procedure builds a library of transition moments of selected one-electron operators.
The library can be used to visualize and analyze the time-evolution of a molecular system, 
represented as a time-dependent linear combination of multi-determinantal wave functions.\\

\begin{figure}[hb]
\centering
\includegraphics{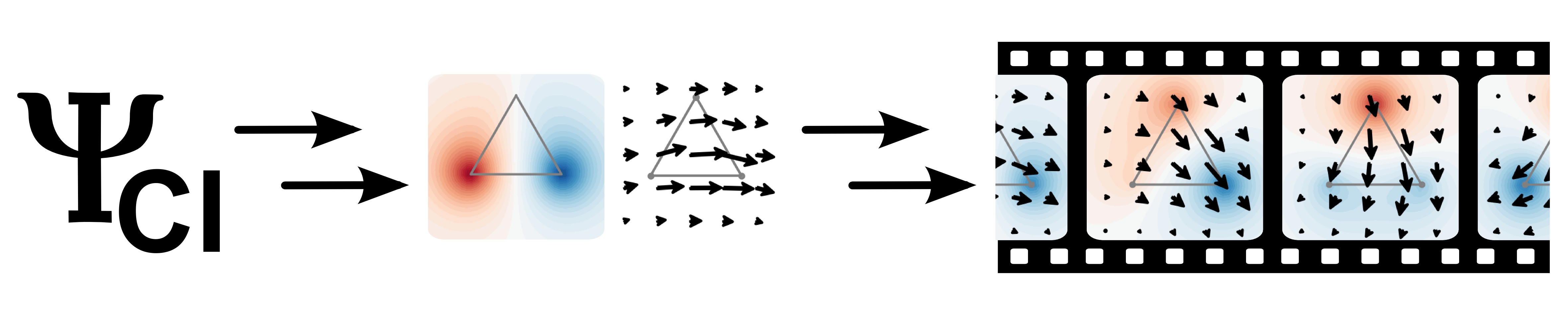}
\label{TOCfig}
\end{figure}

\end{document}